\documentclass[a4paper]{article}
\usepackage{spconf,amsmath,graphicx}
\usepackage{float}
\usepackage{algorithm}
\usepackage{amsfonts}
\usepackage{ifthen}
\usepackage{supertabular}
\usepackage{array}
\usepackage{multirow}
\usepackage{fancyhdr}
\usepackage{hhline}
\usepackage{cite}
\usepackage{fancybox}
\usepackage{ctable}
\usepackage{amssymb}
\usepackage{upgreek}
\usepackage{xcolor}

\usepackage[caption=false]{subfig}

\title{A Momentum Two-gradient Direction Algorithm with Variable Step Size Applied to Solve Practical Output Constraint Issue for Active Noise Control}
\name{Xiaoyi Shen, Dongyuan Shi, Zhengding Luo, Junwei Ji, Woon-Seng Gan}
\address{School of Electrical and Electronic Engineering, Nanyang Technological University, Singapore.\\Email: xiaoyi003@e.ntu.edu.sg}
\begin{document}
\ninept
\maketitle
\begin{abstract}
Active noise control (ANC) has been widely utilized to reduce unwanted environmental noise. The primary objective of ANC is to generate an anti-noise with the same amplitude but the opposite phase of the primary noise using the secondary source. However, the effectiveness of the ANC application is impacted by the speaker's output saturation. This paper proposes a two-gradient direction ANC algorithm with a momentum factor to solve the saturation with faster convergence. In order to make it implemented in real-time, a computation-effective variable step size approach is applied to further reduce the steady-state error brought on by the changing gradient directions. The time constant and step size bound for the momentum two-gradient direction algorithm is analyzed. Simulation results show that the proposed algorithm performs effectively in the time-unvaried and time-varied environment.

\end{abstract}


\vspace{-0.3cm}\section{Introduction}
Active noise control (ANC) is commonly employed to reduce ambient noise~\cite{kuo1996active,elliott1993active, kajikawa2012recent, hansen1999understanding, yang2018frequency,zhang2020active,chang2020active,SHEN2022117300,ShiFeedforward2020,shen2022adaptive,lam2020active,SHEN2021107712,cheer2013practical}. In a single-channel feedforward ANC system, a reference microphone and an error microphone are used to pick up the reference and error signals. The collected signals are used to generate the control signal played back by the secondary source ~\cite{kuo1999active,CHEER2015753,elliott2000signal}.  Output saturation occurs when the maximum output power of the ANC system exceeds the secondary source's driving capacity. The noise reduction performance will degrade once the driving signal of the ANC system surpasses the threshold of output saturation, and the secondary source's inherent nonlinearity could result in the adaptive algorithm divergence throughout the noise reduction process~\cite{costa2001stochastic,kwan2002adaptive}. 

There is a recent interest in adapting the control filter, yet not over-driving the sound level in several practical ANC application~\cite{kuo2004saturation,shi2019practical}.  A bilinear filtered-X least mean square (FXLMS) algorithm is utilized to reduce the distortion of signal saturation~\cite{kuo2005nonlinear}. The bilinear filters employing feedforward and feedback polynomials have the same input-output equation as IIR filters, which accurately model the nonlinear output. Nevertheless, the bilinear algorithm raises the computational requirements when implementing feedforward, feedback, and the cross coefficient for the control filters. A filter bank-based functional link artificial neural network structure~\cite{rout2016particle} is tuned using a particle swarm optimization approach without secondary path training. A rescaling approach is utilized to reduce the output signal and regulate filter coefficients~\cite{qiu2001study,lan2002weight}. The application of the two-gradient (2GD) algorithm ~\cite{shi2019practical} depends on whether the control signal's power exceeds the output power constraint. If the control signal falls within the constraint, the weights of the control filter update using the same method as the FXLMS algorithm. However, if the control signal violates the output constraints, the gradient direction switches, causing the weights to bounce off the boundary and reduce the output power. The  oscillation brought on by the varying gradient increases the steady-state error, and a variable step size strategy is utilized to reduce it. However, the small step size (determined by the variable step size strategy) leads the ANC algorithm to converge slowly, especially in a time-varied environment. In this paper, we apply a momentum mechanism~\cite{shi2020active,roy1990analysis} to the 2GD-FXLMS with variable step size and examine the time constant and step size bounds in two directions for the proposed approach.

The paper is organized as follows. Section~\ref{sec_2} proposes the momentum two-gradient (2GD-Momentum) algorithm with variable step size and analyzes the convergence performance with step size bounds and time constant. Sections~\ref{sec_3}  presents simulation results. Section~\ref{sec_4} concludes this paper.

\section{Proposed method} \label{sec_2}
\begin{figure}[!t]
    \centering
    \includegraphics[width =8cm]{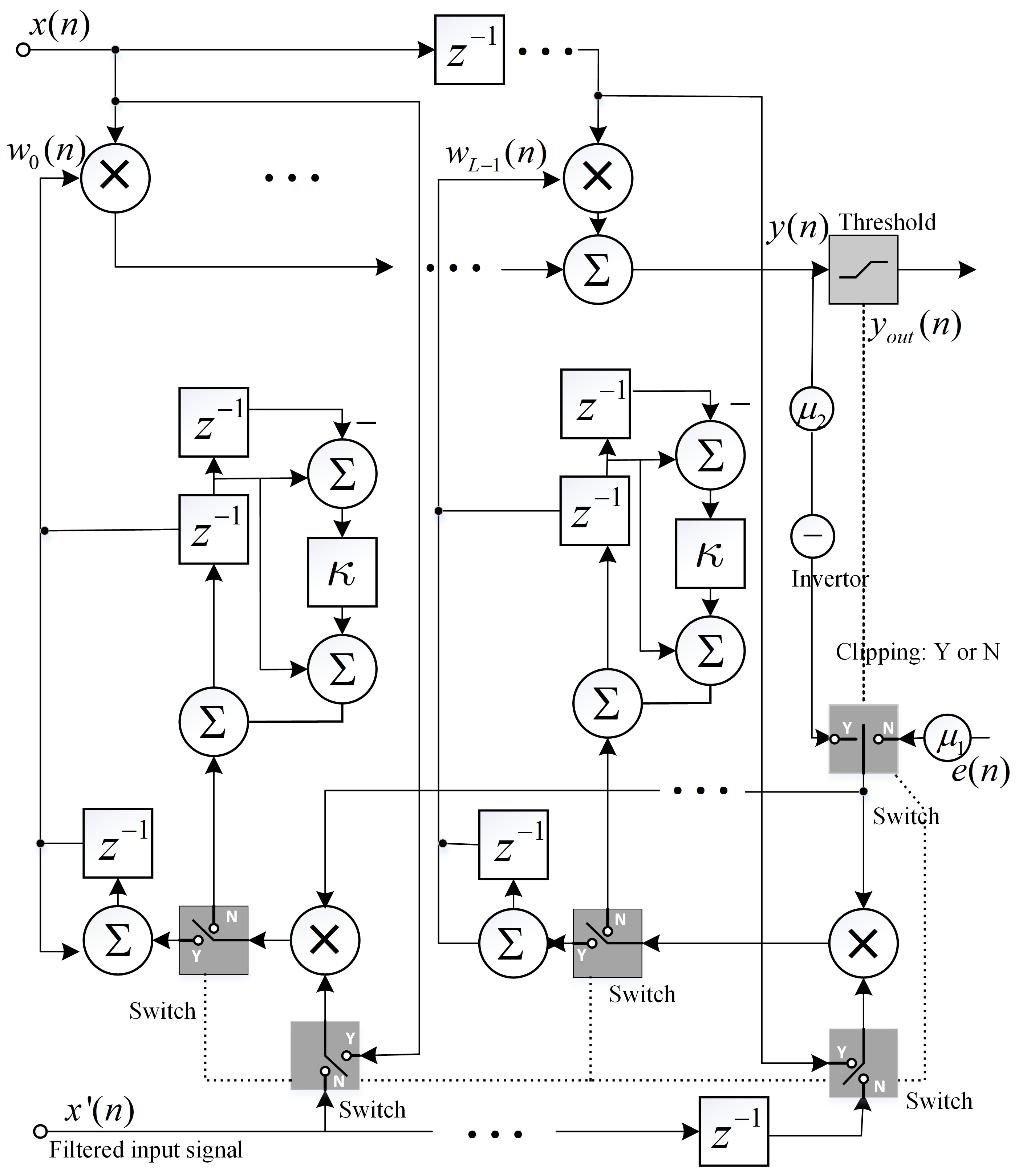}
    \caption{The block diagram of momentum two-gradient algorithm with variable step-size. }
    \label{fig_6}
\end{figure}

In this section, the 2GD-Momentum technique with variable step size for output-constrained ANC is proposed. Fig.\ref{fig_6} shows the block diagram of the proposed method implemented with a feedforward ANC structure. The error signal $e(n)$ is the acoustic summation of the disturbance and the anti-noise, which is formed by the control signal $y(n)$ passing through the secondary path $s(n)$. 

\begin{equation}
    e(n) = d(n) - y(n)\ast s(n),
\end{equation}
 where $\ast$ denotes linear convolution. The control signal $y(n)$ is the output of control filter  $W(z)$: 
 \begin{equation}
     y(n) = \mathbf{w}(n) \mathbf{x}(n),
 \end{equation}
where $\mathbf{w}(n)$ is the coefficients of the feedforward control filter $W(z)$ with $L_f$ taps. $\mathbf{x}(n)$  denotes the reference signal vector.
To solve the output-saturation issue caused by the secondary source, the 2GD-FXLMS algorithm has been developed previously ~\cite{shi2020active}. The maximum output power of the secondary source is defined at $\rho^2$, and the coefficients of the control filter are updated as 
\begin{equation}\label{eq_1}
\mathbf{w}(n+1)=
 \begin{cases} 
\mathbf{w}(n) +\mu_1(n)e(n)\mathbf{x}'(n),  & \mathbb{E}\left[y^2(n)\right]\leq\rho^2 \\
\mathbf{w}(n) - \mu_2y(n)\mathbf{x}(n), & \mathbb{E}\left[y^2(n)\right]>\rho^2
\end{cases}
\end{equation}
 where $\mu_1(n)$ and $\mu_2$ represent the step sizes. $\mathbf{x}'(n)$ denotes the reference signal $\mathbf{x}(n)$ filtered by the secondary path estimated as $\hat{s}(n)$
 \begin{equation}
     \mathbf{x}'(n) = \mathbf{x} (n) \ast \hat{s}(n).
 \end{equation}
However, the unappropriated step size selection will cause serious weight oscillation when alternative updating control filter. To alleviate the weight error, a variable step size strategy is applied
\begin{equation}\label{eq_2}
    \mu_1(n+1) =  
    \begin{cases}
    \mu_1(n), & \mathbb{E}\left[y^2(n)\right]\leq\rho^2\\
    \min\left[ \gamma \mu_1(n), \mu_{\text{min}}\right], &\mathbb{E}\left[y^2(n)\right]>\rho^2
    \end{cases}
\end{equation}
where $\gamma$ and $\mu_{\text{min}}$ represent the common ratio and smallest step size. Once $\mathbb{E}\left[y^2(n)\right]>\rho^2$, the step size changes to $\mu_2$, which is determined by $\mu_1(n)$
 \begin{equation} \label{eq_20}
     \mu_2 = \varsigma \mu_1(n),
 \end{equation}
 where $\varsigma$ denotes the Lagrangian factor obtained from
 \begin{equation}
    \varsigma = G_s(\eta-1),
 \end{equation}
 where $G_s$ denotes the power gain of the secondary path, which is modeled offline.  $\eta$ is the system non-linearity given by
 \begin{equation}
      \eta^2 = \max \left(\frac{\sigma^2_\mathrm{d}}{G_\mathrm{s}\rho^2},1\right),
 \end{equation}
where $\delta^2_d$ represents the power of disturbance $d(n)$, and $\max(\cdot)$ returns the maximum value of the inputs.

In this situation, the algorithm would gradually reduce the power of the control signal until it falls within the output constraint. However, the reduced step size will undoubtedly affect the convergence behavior when the algorithm updates along the first gradient direction, which would become more serious when the acoustic environment changes. Therefore, we bring the momentum factor in the weight updating equation to accelerate the convergence:

\begin{equation}\label{eq_3}
\mathbf{w}(n+1)=
 \begin{cases} 
 \mathbf{w(n)}+  \boldsymbol{\zeta}(n),  & \mathbb{E}\left[y^2(n)\right]\leq\rho^2 \\
\mathbf{w}(n) -\mu_2y(n)\mathbf{x}(n), & \mathbb{E}\left[y^2(n)\right]>\rho^2
\end{cases}
\end{equation}
where $\boldsymbol{\zeta}(n)$ denotes the momentum factor
\begin{equation}
      \boldsymbol{\zeta}(n+1) = \kappa   \boldsymbol{\zeta}(n) + \mu_1(n) e(n)\mathbf{x}'(n),
\end{equation}
where $\kappa$ is the forgetting factor. By involving the momentum technique, the accumulation effect can somehow compensate for the reduced gradient value caused by the variable step size, improving the algorithm's convergence speed.  

\subsection{Convergence analysis}
In order to investigate the convergence performance of the proposed 2GD-Momentum, we define the weight error vector as

\begin{equation}
   \mathbf{v}(n) = \mathbf{w}_o - \mathbf{w}(n),
\end{equation}
where $\mathbf{w}_o$ represents the optimal control filter. 
If the desired output power falls within the output constraint ($\mathbb{E}\left[y^2(n)\right]\leq\rho^2$), the proposed algorithm will achieve the same optimal solution as FXLMS~\cite{kuo1996active}:
\begin{equation}
    \mathbf{w}_o =\mathbf{P}_{dx'}\mathbf{R}^{-1}_{x'},
\end{equation}
where $\mathbf{P}_{dx^\prime}$ denotes the cross power spectrum of $d(n)$ and $x'(n)$, and $\mathbf{R}_{x^\prime}$ represents the power spectrum of $x'(n)$. 
While the desired output power exceeds the constraint, the power of the control signal is limited to the maximum value ($\mathbb{E}\left[y^2(n)\right] = \rho^2$), and the algorithm achieves a sub-optimal control filter as its optimal solution under the output constraint~\cite{SHI2019651}:   
\begin{equation}
    \mathbf{w}_o^{\text{sub}} =\mathbf{P}_{dx}({\varsigma \mathbf{R}_{x}+\mathbf{R}_{x'}})^{-1} .
\end{equation}

\subsubsection{Case 1: $\mathbb{E}\left[y^2(n)\right] \leq \rho^2$}
In this case, the weight error vector is derived to 

\begin{equation} \label{eq_4}
\begin{split}
& \mathbf{v}(n+1)-\mathbf{v}(n) =
    \kappa \left[ \mathbf{v}(n)-\mathbf{v}(n-1) \right] \\&-\mu_1(n) e_o(n)\mathbf{x}'(n)-\mu_1(n)\mathbf{x}'(n)\mathbf{x}'^{\mathrm{T}}(n)\mathbf{v}(n),  
\end{split}
\end{equation}
where 

\begin{equation}
    e_o(n) = d(n) - \mathbf{w}^\mathrm{T}_o \mathbf{x}(n)\ast s(n).
\end{equation}
By applying Kushner's Direct-average method \cite{haykin2002adaptive} into \eqref{eq_4}
\begin{equation}\label{eq_6}
\begin{split}
    &\bar{\mathbf{v}}(n+1)-\bar{\mathbf{v}}(n)\\&=
  \kappa \left[\bar{\mathbf{v}}(n)-\bar{\mathbf{v}}(n-1) \right] -\mu_1(n)\mathbf{R_{x'}}\bar{\mathbf{v}}(n).   
\end{split}
\end{equation}
Here we define
\begin{equation}\label{eq_8}
    \mathbf{\acute{v}}(n) = \mathbf{Q}^{\mathrm{T}}\bar{\mathbf{v}}(n),
\end{equation}
$\mathbf{Q}$ is the unitary matrix constitute with the eigenvectors of $\mathbf{R}_{x'}$. Therefore, it can be found that  
\begin{equation}
 \mathbf{Q}^{\mathrm{T}}\mathbf{R}_{x'}\mathbf{Q} = \mathbf{\Lambda}_{x'},
\end{equation}
and
\begin{equation}
 \mathbf{Q}^{\mathrm{T}}\mathbf{Q} = \mathbf{I},
\end{equation}
where $\mathbf{I}$ denotes the identity matrix. Hence, \eqref{eq_6}  can be rewritten as 
\begin{equation}\label{eq_9}
\begin{split}
     &\mathbf{\acute{v}}(n+1) - \mathbf{\acute{v}}(n) \\ 
     &=\kappa\left[\mathbf{\acute{v}}(n) - \mathbf{\acute{v}}(n-1)\right]-\mu_1(n)\mathbf{\Lambda}_{x'} \mathbf{\acute{v}}(n).
\end{split}
\end{equation}
To further rewritten \eqref{eq_9}
\begin{equation}
   \begin{bmatrix} 
   \acute{v}_k(n+1)\\\acute{v}_k(n)
   \end{bmatrix} =   \begin{bmatrix} (1+\kappa)\mathbf{I}-\mu_1(n)\lambda_k  &-\kappa\mathbf{I}\\
   \mathbf{I} &0\end{bmatrix} \begin{bmatrix} 
   \acute{v}_k(n)\\\acute{v}_k(n-1)
   \end{bmatrix}, 
\end{equation}
where $\lambda_k$ denotes the $k$th ($k=1,2,\cdots,L_f$) eigenvalue of $\mathbf{R}_{x'}$. The stability of the above equation can be instigated through the roots of $\beta_i$ of the determinant\cite{roy1990analysis,kailath1980linear}
\begin{equation}\label{eq_10}
     det \begin{bmatrix} (1+\kappa-\beta_i)\mathbf{I}-\mu_1(n)\lambda_k &-\kappa\mathbf{I}\\
   \mathbf{I} & -\beta_i \mathbf{I} \end{bmatrix} =0.
\end{equation}
The necessary and sufficient condition of the stability should be $\begin{vmatrix}\beta_i \end{vmatrix} <1$. A typical quadratic form is derived from \eqref{eq_10} as
 \begin{equation} \label{eq_22}
     \beta_i^2-\beta_i\left[1+\kappa-\mu_1(n)\lambda_i\right]+\kappa =0.
 \end{equation}
From the above equation, the stability condition for the algorithm is obtained as 
\begin{equation}
    \begin{cases}
 |\kappa | < 1,\\
  0<\mu_1(n)<\frac{1+\kappa}{\lambda_{\text{max}}},
    \end{cases}
\end{equation}
where $\lambda_{\text{max}}$ denotes the largest eigenvalue of $\mathbf{R}_{x'}$ . 
\subsubsection{Case 2: $\mathbb{E}\left[y^2(n)\right] > \rho^2$}
In this case, we aims to minimize $\mathbb{E}\left[y^2(n)\right]$, and hence, the optimal weight is $\mathbf{w_o}=\mathbf{0}$ in the ideal case (if no power constraint applied). 

The weight error vector in Case 2 is determined by
\begin{equation}\label{eq_5}
\begin{split}
  \mathbf{v}(n+1) =\left[ \mathbf{I}-\mu_2\mathbf{x}(n)\mathbf{x}(n)\right]\mathbf{v}(n).
\end{split}
\end{equation}
By applying Kushner's Direct-average method into
\eqref{eq_5}
\begin{equation}\label{eq_7}
  \mathbf{\bar{v}}(n+1) =\left[ \mathbf{I}-\mu_2\mathbf{R}_{x}\right]\mathbf{\bar{v}}(n). 
\end{equation}
Simplify and substitute $\mathbf{\acute{v}}(0)$ into \eqref{eq_7}
\begin{equation}
\begin{split}
   \acute{v}_k(n+1) =\left[ 1-\mu_2\lambda_k\right]^{n}\acute{v}_k(0).
\end{split}
\end{equation}
To get $\mathbf{\acute{v}}(n+1) =0$ at the steady-state, we need to make sure $\left[ 1-\mu_2\lambda_k\right]^{n} \rightarrow 0$, hence , combine \eqref{eq_20} it can get
\begin{equation} \label{eq_24}
     0<\varsigma \mu_1 (n)<\frac{1}{\lambda_{\text{max}}}.
\end{equation}
\vspace{-0.5cm}\subsection{Time constant}
The time constant of the updating equation based on the first gradient is derived in this section. The time constant is related to~\cite{widrow1977stationary}
\begin{equation}\label{eq_23}
    1-\frac{1}{\tau_i} \simeq \begin{vmatrix}\beta_i \end{vmatrix} .
\end{equation}
By solving \eqref{eq_22} and substituting the solution into \eqref{eq_23}, the time constant can be obtained from 
\begin{equation}
    \tau_i \simeq (\frac{1+\kappa}{1+2\kappa}) \frac{1}{2\mu_1(n)\lambda_{\text{min}}}.
\end{equation}
If the momentum factor is set to $0<\kappa<1$, it can be inferred from \eqref{eq_20} and \eqref{eq_24} that the step size is greater than the FXLMS. Therefore, it can be concluded that the momentum factor decreases the time constant value and thereby accelerates the ANC's convergence process.

\vspace{-0.3cm}\section{Simulation Results}\label{sec_3}
The effectiveness of the proposed method is evaluated through simulations. In the first simulation, the saturation effect brought on by the output constraints is verified. The weight fluctuations of the control filter in a static and varying environment are then demonstrated in the subsequent simulations.
\subsection{Saturation effect }
\begin{figure}
    \centering
    \includegraphics[width=8cm]{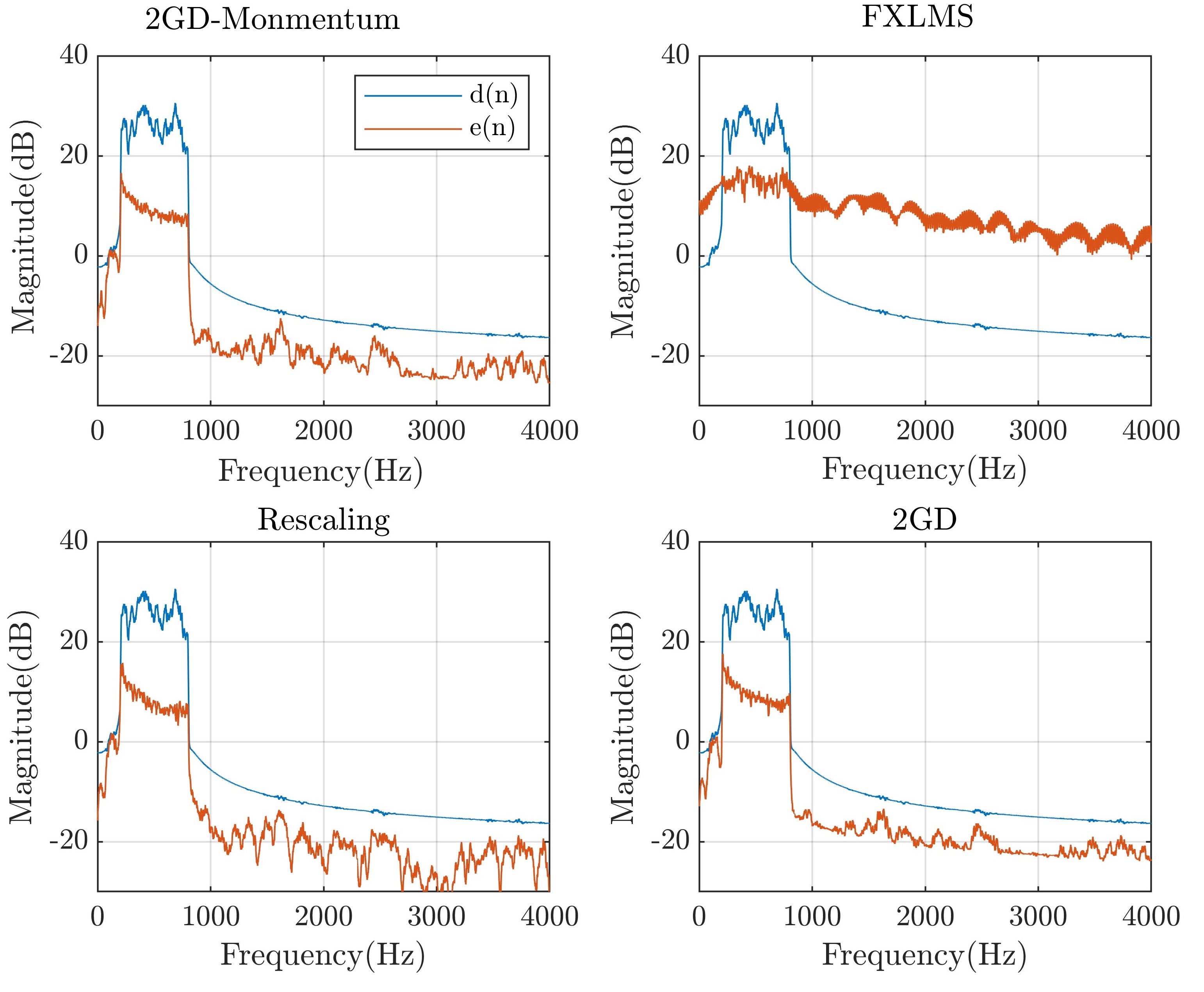}
    \caption{Power spectrum of error signals in momentum two-gradient direction (2GD-Momentum) with variable step size, FXLMS algorithm, rescaling and two-gradient direction (2GD) when the output power exceeds the constraint.  }
    \label{fig_1}
\end{figure}

 In order to simulate the saturation effect, a clipping function is applied as follows: 
\begin{equation}
  y(n)=  \begin{cases}
    y(n),  & \mbox{if } y(n) < y_T \\
    y_T.   & \mbox{if } y(n) \ge y_T
    \end{cases}
\end{equation}
In the simulation, the output constraint $y_T$ is set to $1$. The initial step size is $\mu_1(0) = 0.00001$, $\mu_{\text{min}} = 0.000001$, and the parameters used to generate the variable step sizes are $\varsigma =0.85$, $\gamma =0.9$. The forgetting factor $\kappa =0.99$. The control filter's length is $512$ taps, and the sampling frequency is $16$ kHz. The simulation's primary and secondary paths are measured from an air duct. As a comparison, the FXLMS algorithm, rescaling algorithm~\cite{lan2002weight}, and two-gradient direction (2GD) algorithm~\cite{SHI2019651} are tested. A broadband noise with $200-800$ Hz is applied as the primary noise, and the power spectrum of the error signal is shown in Fig.\ref{fig_1}. It can be figured out that with the output constraint, the proposed method, the rescaling algorithm, and the 2GD algorithm attenuate the noise during $200-800$ Hz. In contrast, the FXLMS algorithm without output constraint has the boosting frequency components outside the $200-800$ Hz band.
\subsection{Noise cancellation in a static environment}
\begin{figure}[!t]
    \centering
    \includegraphics[width=7cm]{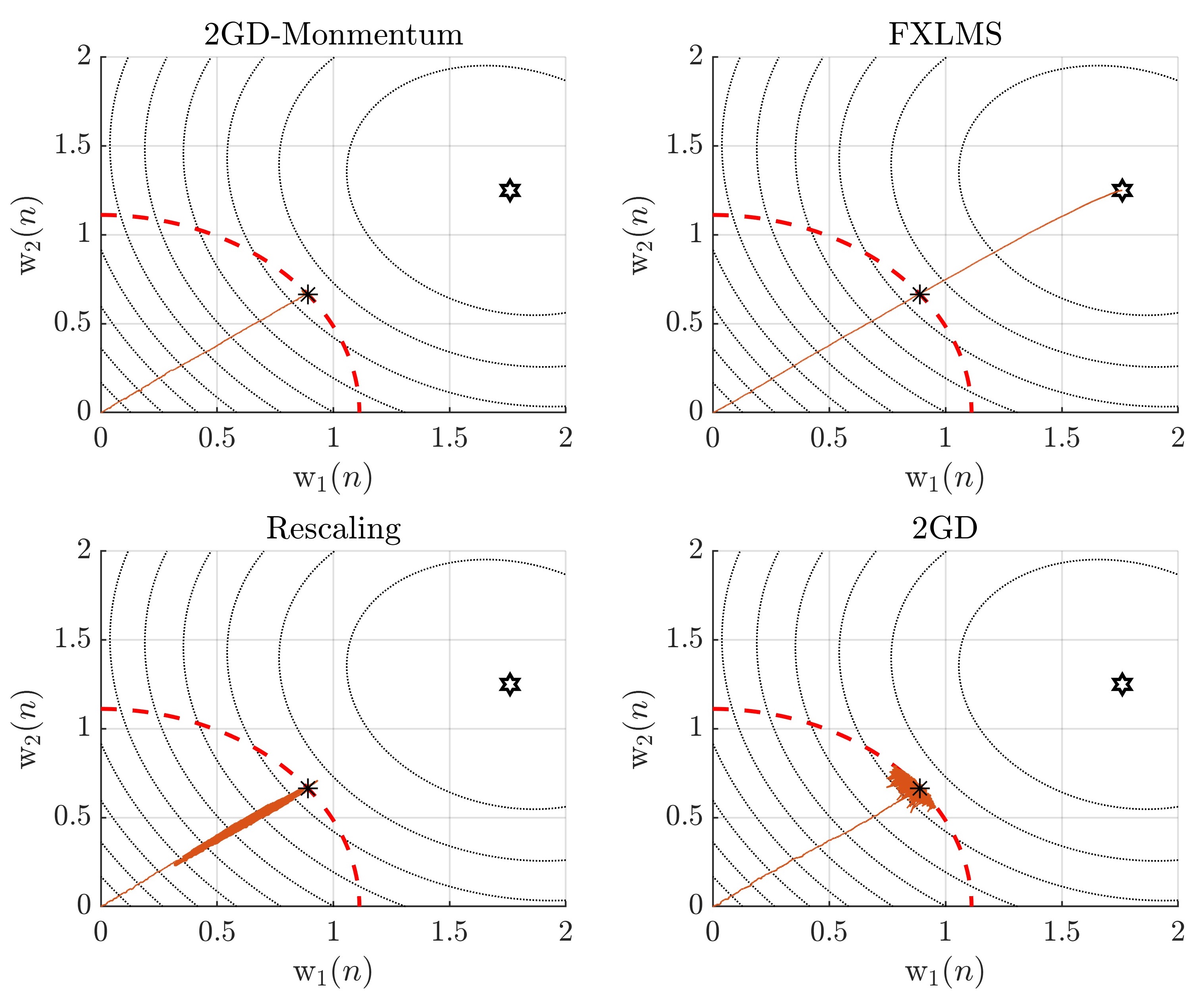}
    \caption{The convergence path in a static environment during the noise reduction process. }
    \label{fig_4}
\end{figure}

\begin{figure}[!t]
    \centering
    \includegraphics[width=7cm]{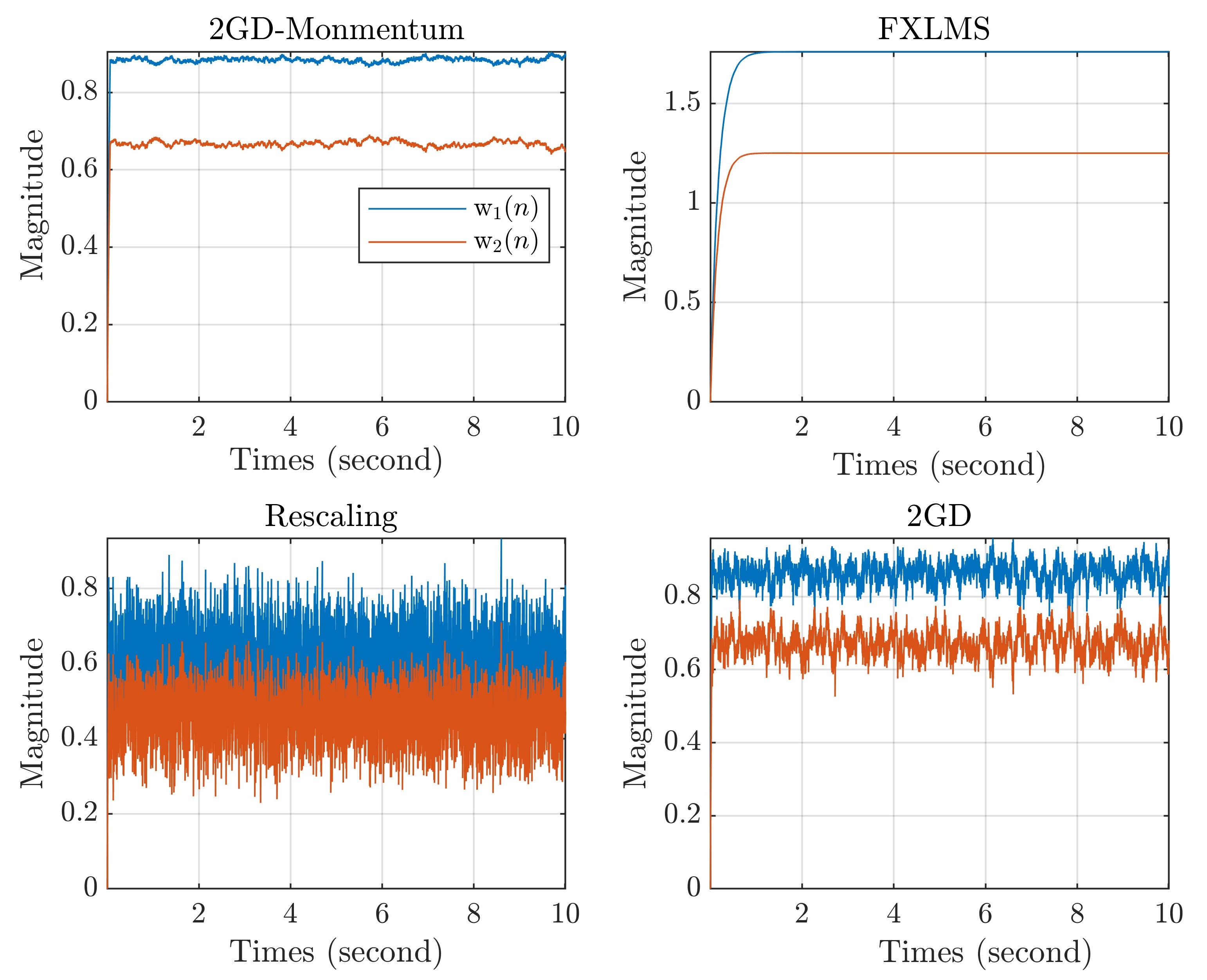}
    \caption{The weight variations in a static environment during the noise reduction. }
    \label{fig_5}
\end{figure}

The simulation evaluates the control filter's weight in a time-unvaried environment. The white Gaussian noise is used as the primary noise. To visualize the variation of the weights, we utilized a two-weight control filter, 
and the primary path and secondary path are set at $\left[1.76~1.25 \right]$ and $\left[0.13~0.87 \right]$. The optimal value without weight constraints are derived as $\mathbf{w}_o=\left[1.76~1.25 \right]$, and the sub-optimal weights is $\mathbf{w}^{sub}_o=\left[0.89~0.66 \right]$ , which is presented as $\star$ and $*$ in Fig.\ref{fig_4}.  The weight constraint is shown in the red dashed line. As seen in Fig.\ref{fig_4}, the proposed 2GD-Momentum reaches the sub-optimal value when scaled and no longer increases to the optimal weight. The FXLMS algorithm simultaneously bypasses the boundary and reaches the optimal weight that is out of the constraints. The proposed method reduces the vibration effect of weights present in the rescaling and 2GD algorithms when they achieve the constraint.  As Fig.\ref{fig_5} shows, the proposed method lowers the oscillating effect when the control filter reaches sub-optimal weight.
\vspace{-0.3cm}\subsection{Noise cancellation in a varying environment}
\begin{figure}[!t]
    \centering
    \includegraphics[width=7cm]{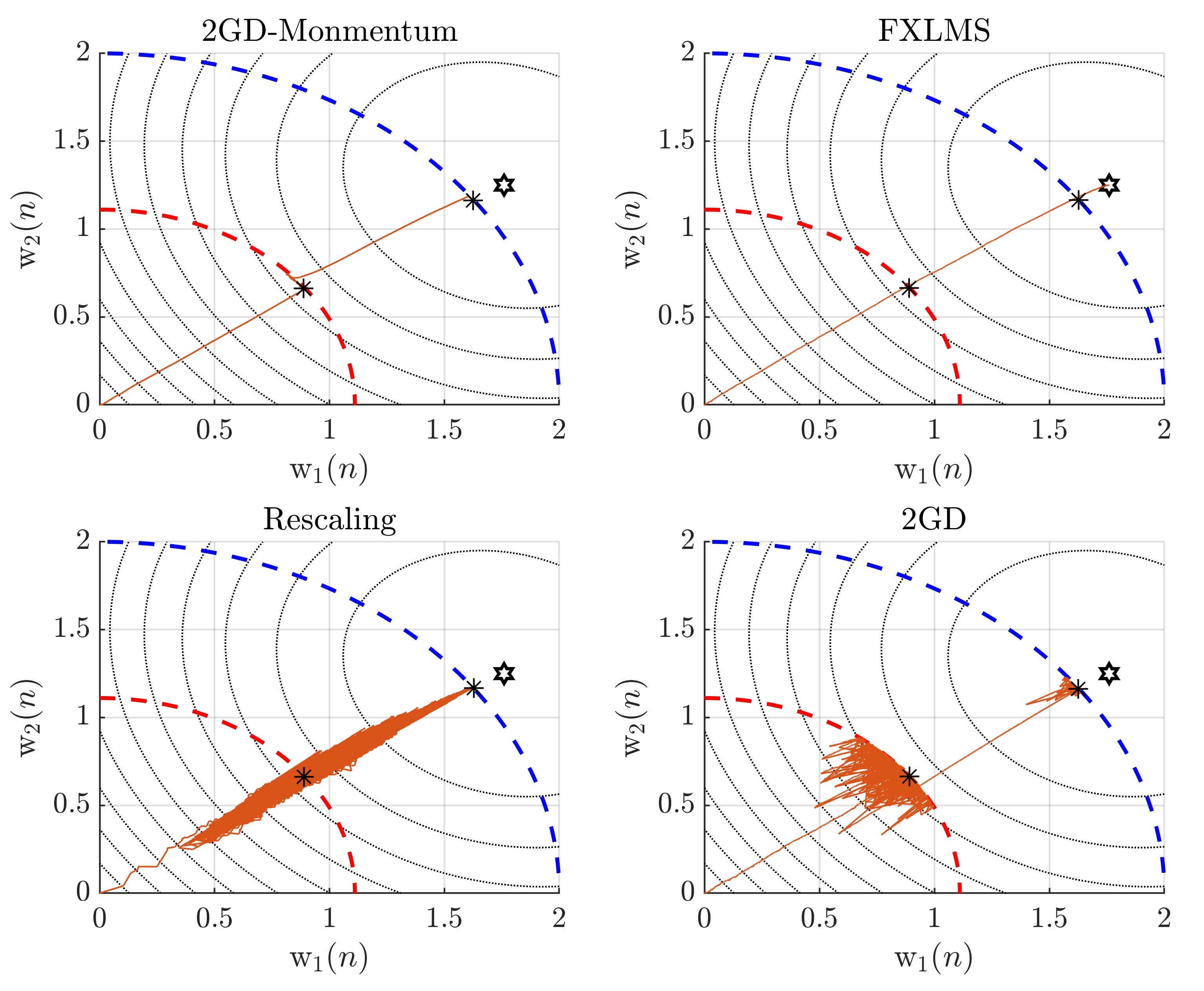}
    \caption{The convergence path when the primary path changes during the noise reduction. }
    \label{fig_2}
\end{figure}

\begin{figure}[!t]
    \centering
    \includegraphics[width=7cm]{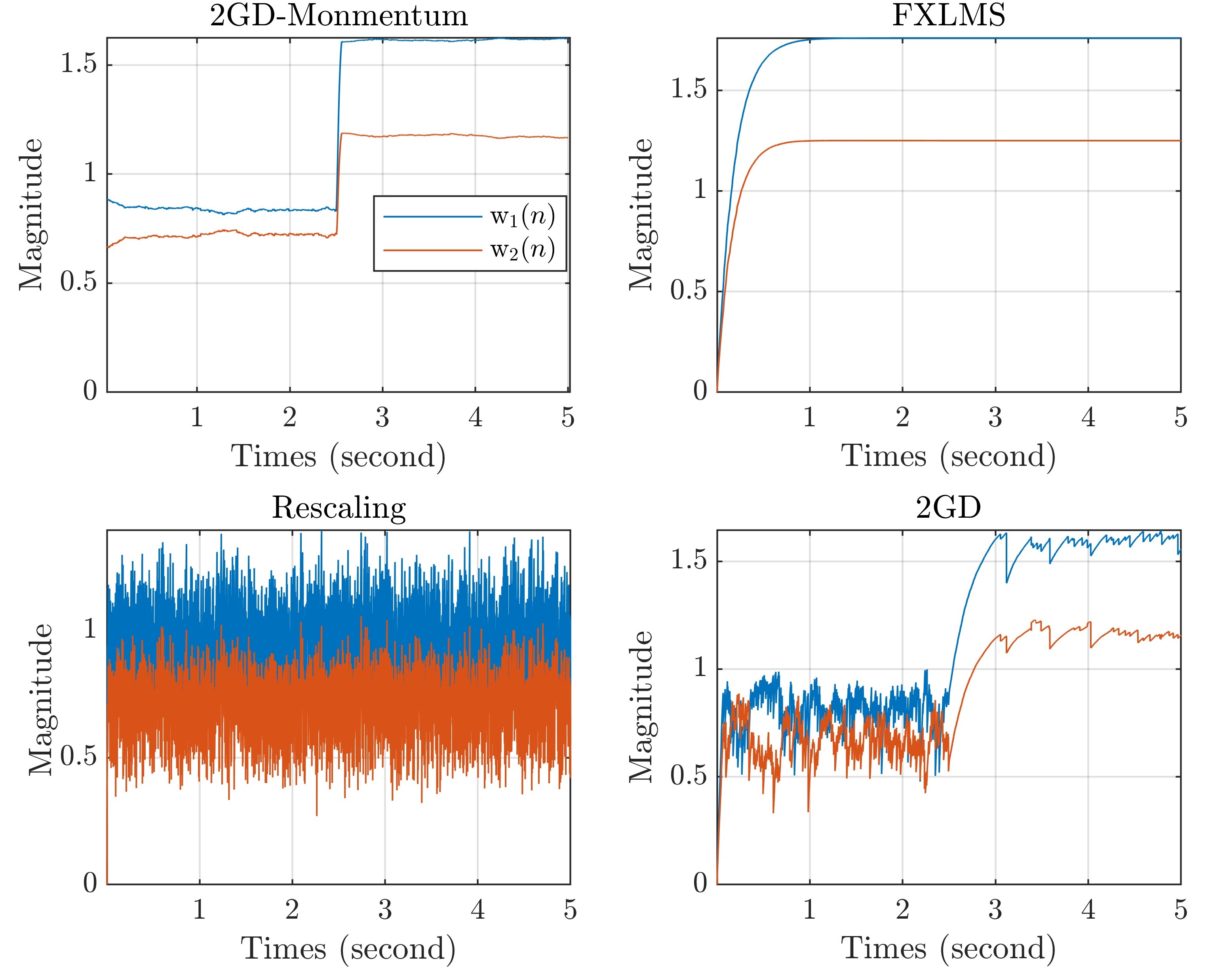}
    \caption{The weight variations when the primary path changes during the noise reduction. }
    \label{fig_3}
\end{figure}

The simulation evaluates the control filter's weight in a time-varied environment. The simulation's configuration is the same as in the last simulation before the environment changed. During the noise reduction process, the acoustic environment changes, and the second sub-optimal value changes from $\mathbf{w}^{sub}_o=\left[0.89~0.66 \right]$ to  $\mathbf{w}^{sub}_o=\left[1.63~1.17 \right]$, as shown in Fig.\ref{fig_2}.  The new weight bound is shown as the blue dashed line in Fig.\ref{fig_2}. Although the rescaling and 2GD algorithm also achieve the second sub-optimal value, the proposed 2GD-Momentum method has a faster convergence speed without weight vibration, as shown in Fig.\ref{fig_3}.
\vspace{-0.2cm}\section{Conclusion}\label{sec_4}
The two-gradient FXLMS algorithm was a practical solution to overcome the output saturation issue in adaptive ANC systems and increase system stability. However, its weight oscillation caused by gradient switching progress influenced the noise reduction performance. This paper applied the momentum two-gradient direction algorithm with variable step size to output-constrained ANC. The momentum factor, along with the variable step size approach, was added to reduce the weight oscillation brought on by the two-gradient method and speed up convergence in a time-varied environment. The analysis of the two-gradient method's step size bounds and time constant demonstrated its accessibility and enhanced convergence capabilities. According to the simulation results, the proposed approach had an advantage in terms of convergence speed and a decreased vibration effect on weight changes for the two-gradient step size.
 
\vspace{-0.3cm}\section{Acknowledgments}
This research/work is supported by the Singapore Ministry of National Development and National Research Foundation under the Cities of Tomorrow R$\&$D Program: COT-V4-2019-1.

\bibliographystyle{IEEEbib}
\bibliography{a}
\end{document}